\documentstyle[aps,pre,epsf,floats]{revtex}
\def\figuresize{\ifpreprintsty 12cm \else 8cm \fi}

\begin{document}

\ifpreprintsty \else
\twocolumn[\hsize\textwidth\columnwidth\hsize\csname@twocolumnfalse%
\endcsname \fi

\draft
\title{Differences between regular
   and random order of updates in damage spreading simulations}

\author{Thomas Vojta and Michael Schreiber}
\address{Institut f\"ur Physik, Technische Universit\"at, D-09107 Chemnitz, Germany}
\date{version Aug 12, printed \today}
\maketitle

\begin{abstract}
We investigate the spreading of damage in the three-dimensional Ising model 
by means of large-scale Monte-Carlo simulations. Within the Glauber dynamics we use 
different rules for the order in which the sites are updated. We find that the stationary
damage values and the spreading temperature are {\em different} for {\em different update order}.
In particular, random update order leads to larger damage and a lower spreading temperature
than regular order. Consequently, damage spreading in the Ising model is non-universal not only
with respect to different update algorithms (e.g. Glauber vs. heat-bath dynamics) as already 
known, but even with respect to the order of sites.
\end{abstract}

\ifpreprintsty \else
] \fi              


Damage spreading (DS) investigates how a small perturbation in a cooperative system 
changes during the time evolution. It was first studied in theoretical biology \cite{kauffman}
in the context of the genetical evolution. Later the DS concept found its way into the physics
of cooperative systems \cite{creutz,derrida1,stanley}.
In order to study  DS  two replicas of the system are considered which evolve stochastically 
under the same noise realization (i.e. the same random numbers are used in a  
Monte-Carlo procedure).  The difference in the microscopic configurations of the two replicas 
constitutes the "damage". Depending on the Hamiltonian, the dynamic rules, and the external
parameters a small initial damage will either spread or heal with time (or remain finite
in a finite spatial region). 
This behavior distinguishes chaotic or regular phases.
It was realized early on that the properties of DS depend sensitively on the update rule employed 
in the Monte-Carlo procedure. For instance, in the Ising model with Glauber dynamics
\cite{stanley} 
the damage heals at low temperatures and spreads at temperatures above a certain
spreading temperature $T_s$. In contrast, 
the Ising model with heat-bath dynamics \cite{derrida1} shows qualitatively different behavior: 
the damage heals at high 
temperatures but it may freeze at low temperatures. Thus, DS appears to be uniquely 
defined only if one specifies the Hamiltonian {\em and} the dynamic rule. (Note that is was suggested
\cite{hinrichsen} to obtain an unambiguous definition of DS for a particular model by considering all 
possible dynamic rules which are consistent with the physics of a single replica.)
The differences between Glauber and heat-bath dynamics which can be traced back to 
different use of  the random numbers in the update rules \cite{rn} can be understood
already on the basis of a mean-field theory for DS \cite{vojta1}. 

In addition to this dependence of DS on the update rule (i.e. the way the random numbers are
used in the simulation) it was also found \cite{nobre} that in some systems DS can be completely different 
for parallel instead of sequential updates of the lattice sites. This is not too surprising since
even the equilibrium probability distributions are different for parallel and sequential updates.

In this Brief Report we investigate the dependence of DS on another detail of the Monte-Carlo 
procedure employed in the simulation, viz. the order of sites within a sequential update scheme.
In general, different update schemes define different dynamical systems which 
will show different dynamical behavior. While all update schemes which differ only in the 
order of the sites will lead to the same stationary (equilibrium) state for a {\em single} replica (thanks to detailed
balance) the same is not a priori true for DS which is a non-equilibrium phenomenon.
To the best of our knowledge the question whether the stationary state of  DS (i.e the stationary state of  the 
{\em pair} of replicas) does depend on the site order in the update scheme has not been
investigated before \cite{cellu}. Most of the published work on DS in the Ising model  seems to (implicitly) assume 
that at least the stationary damage (and thus the spreading temperature) do not depend on the site
order. In this Brief Report we provide numerical evidence that this assumption is mistaken.

We have studied DS in the Glauber Ising model on a cubic lattice with $N=L^3$ sites.  
The Hamiltonian is given by
\begin{equation}
H = - \frac 12 \sum _{  ij } J_{ij }~S_i S_j
\label{eq:hamiltonian}
\end{equation}
where $S_i= \pm 1$ is the Ising variable at site $i$, $J_{ij}$ is the exchange energy 
which we take to be one for nearest neighbor sites and zero otherwise. The Glauber dynamics 
is given by the stochastic map 
\begin{equation}
S_i (t+1) = {\rm sign} \left\{ v[h_i(t)] 
- \frac 1 2 +S_i(t) \biggl[ \xi_i(t) - \frac 1 2 \biggr] \right\}
\end{equation}
with the transition probability
\begin{equation}
v(h) = {e^{h/T}/ ({e^{h/T}+ e^{-h/T}}}).
\end{equation}
Here $h_i(t)=\sum_j J_{ij} S_j(t)$ 
is the local magnetic field at site $i$ and (discretized)
time $t$, $T$ denotes the temperature and
$\xi_i(t) \in [0,1)$ is a random number which is identical
for the two copies of the system  considered in a DS
simulation. As in any DS simulation the central quantity
studied is the Hamming distance (damage) $D$ as a function of
time $t$,
\begin{equation}
D(t)= \frac 1 {2N} \sum_i |S_i^{(1)}(t) - S_i^{(2)}(t)|
\end{equation}
where the upper index of the spin variable distinguishes the two replicas.

DS in the Glauber Ising model has been intensively investigated 
both numerically \cite{stanley,costa,caer,grassberger,wang} and using 
an effective field theory \cite{vojta1,vojta2}. The most precise estimate of the
spreading temperature $T_s$  (above which the Hamming distance remains
finite in the long-time limit) in three dimensions 
was obtained in Ref. \cite{grassberger} for systems with
up to $309 \times 309 \times 310$ sites using helical boundary conditions and a
checkerboard update scheme. The result was a spreading temperature of 
$T_s/T_c=0.9225 \pm 0.0005 ~ (T_s=4.162)$ where $T_c= 4.5115$ is  the equilibrium 
critical temperature of the ferromagnetic phase transition (all temperatures are measured 
in units of the nearest-neighbor interaction).

We have carried out extensive DS simulations for systems with up to $N=101^3$ sites 
with periodic and helical boundary conditions giving
both the time evolution of the damage and its asymptotic stationary value. Different update
sequences have been used: typewriter (regularly going from one site to the next), 
checkerboard (regularly going from one site to its next nearest neighbor, effectively updating first 
one sublattice, then the other), and three different types of random sequences. For the first random
sequence the site to be updated is chosen independently for each time step. 
In the second random scheme
each site is updated exactly once during each sweep 
(a sweep consists of $N$ Monte-Carlo updates), 
but the (random) order is different from sweep to
sweep. In the third random procedure, we use identical (random) order in all sweeps.

\begin{figure}
  \epsfxsize=\figuresize
  \centerline{\epsffile{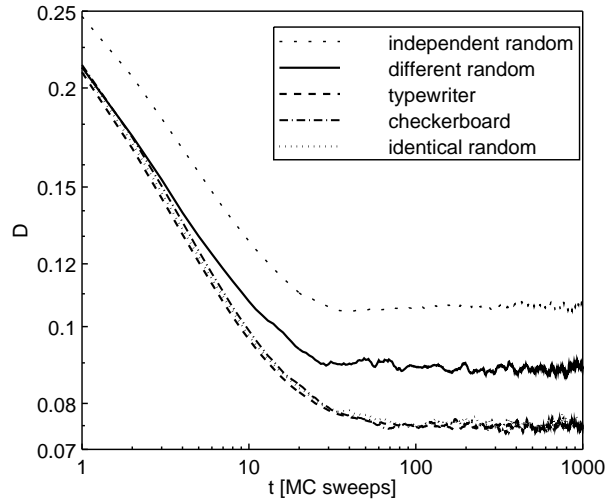}}
  \caption{Time evolution of the damage for a $27^3$ system at temperature $T=4.25$.
     The two copies were prepared independently with an initial magnetization of  $m_0=0.6$ which
     corresponds to an initial damage $D_0=(1-m_0^2)/2=0.32$. The curves
     represent averages over 400 noise realizations.}
  \label{fig:mess}
\end{figure}

In Fig.\ \ref{fig:mess} we show an example for the time evolution of the damage averaged
over 400 runs with different noise realizations. The temperature $T=4.25$ is slightly below
$T_c=4.5115$.
The figure shows that not only the approach to the stationary state but also the stationary damage 
itself depend on the order of sites in the update process. 

The short-time behavior is comparatively easy to understand: If the sites to be updated are chosen 
independently some sites will be updated twice or even several times while some will
not be updated at all during the first sweep through the lattice. In contrast, for all other
update sequences each site is updated exactly once during each Monte-Carlo sweep.
Now, in the example in Fig.\ \ref{fig:mess} the initial damage is higher than its stationary value.
Thus, the damage has to be reduced during the first few sweeps. However, if some sites
are not updated at all, their damage cannot heal and consequently the case of independently
chosen sites leads to slower decrease of the damage within the first few sweeps.
In accordance with this explanation
Fig.\ \ref{fig:mess} shows that after the first sweep the damage is identical for all 
sequences that update each site exactly once in each sweep.

Let us now turn to the stationary states. Fig.\ \ref{fig:mess} indicates that 
the stationary state of the {\em pair} of replicas is indeed different for different site order
in contrast to the stationary state of a {\em single} replica which is independent of
the site order as discussed above. A closer inspection of Fig.\ \ref{fig:mess} shows
that the stationary damage for all those
schemes for which the order of sites does not change from sweep to sweep
 (typewriter, checkerboard, and
identical random) is the same within the statistical accuracy. A significantly higher 
stationary damage value is obtained if we use different random sequences but still 
update each site exactly once in each sweep. 
Finally, for a completely uncorrelated sequence the stationary damage value is 
largest. 
Thus, correlations of the site sequences between {\em different sweeps}
appear to influence the stationary state of the {\em pair} of replicas.
We also note that the mean-field theory \cite{vojta1} cannot explain this new dependence of DS on 
the update sequence since within the mean-field theory the problem is reduced to a single-site
problem.

\begin{figure}
  \epsfxsize=\figuresize
  \centerline{\epsffile{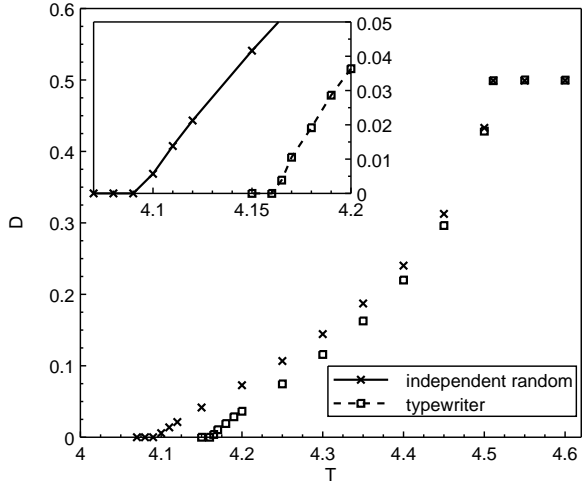}}
  \caption{Temperature dependence of the average stationary damage for typewriter and independent 
     random site sequences. The curves represent averages over 10 runs of a system with $101^3$ sites.
     In each run the damage is averaged over 10000 Monte-Carlo sweeps after a stationary regime has 
     been reached. The inset shows the spreading transition region. The statistical error is smaller than 
     the symbol size in the main figure and 
     approximately given by the symbol size in the inset. }
  \label{fig:ds}
\end{figure}

We have carried out high precision calculations at different temperatures using the various update 
schemes discussed above in order to obtain the temperature dependence of the average stationary 
damage values. In these calculations the two replicas are prepared with a small initial damage. 
The time evolution is monitored and after a stationary regime has been reached
the damage is averaged over a large number ($10^4$) of Monte-Carlo sweeps.
The results for the typewriter and independent random
update schemes are shown in Fig.\ \ref{fig:ds}.
Analogous calculations have been carried out for the other update schemes.
In the paramagnetic phase ($T>T_c$) the average stationary damage value is 0.5 for all update
sequences investigated. In the ferromagnetic phase, however, the results are different. The three
schemes that use the same sequence of sites in all sweeps (typewriter, checkerboard and
identical random) give identical stationary damage averages within the statistical accuracy. For these
schemes we obtain a spreading temperature of  $T_s=4.1625 \pm 0.0050$, i.e., 
$T_s/T_c =0.9225 \pm 0.0010$. This is exactly the value obtained by Grassberger \cite{grassberger}
(using the checkerboard update scheme). 
In contrast, for the independent random sequence the spreading temperature is significantly lower.
We obtain $T_s=4.0950 \pm 0.0050$, i.e. $T_s/T_c=0.9075 \pm 0.0010$. 
The results shown in Fig.\ \ref{fig:ds} also indicate that the critical behavior at the
spreading transition is the same for the update schemes investigated. Since DS in the Glauber Ising model
has two equivalent absorbing states (corresponding to $D=0$ and $D=1$), 
the critical behavior should be in the parity conserving (PC)
universality class \cite{hin_isi}. It has been suggested \cite{sevabstate} that the model with
two absorbing states in three dimensions is already above its upper critical dimension.
It should then have a critical exponent  $\beta=\beta_{mf}=1$, see e.g. Ref.\ \cite{vojta1}
($\beta$ is defined by $D(T) \sim  (T-T_c)^\beta$).
The data in Fig.\ \ref{fig:ds} are roughly consistent with this prediction
for both
update schemes although the inset seems to suggest a slightly smaller exponent.
A systematic investigation of the critical behavior will be published elsewhere \cite{vojta3}.

All the results reported so far have been obtained using periodic boundary conditions.
For comparison we have also investigated the influence of helical boundary conditions. Within 
the statistical accuracy the results of both boundary conditions are the same. 

Furthermore,
we have also checked whether the choice of the random number generator does play any 
role. Three very different random number generators have been used in the simulations:
a combined linear congruential generator (RAN2 from Ref. \cite{recipes}), a very simple 
linear feedback shift register generator (R250, see Ref. \cite{R250}) and a state-of-the-art
combined linear feedback shift register generator (LFSR113 from Ref. \cite{lfsr113}).
All random number generators lead to the same results in our DS simulations. From this 
we exclude any errors due to poor random numbers.

To summarize, we have studied the dependence of damage spreading in the three-dimensional
Glauber Ising model on the order of the sites in the Monte-Carlo update scheme. By using five different
update schemes we have provided numerical evidence that 
the stationary damage and thus the spreading temperature are different for different site order.
For all schemes which use the same site sequences in each sweep (typewriter, checkerboard,
identical random) we have obtained a spreading temperature of $T_s/T_c =0.9225 \pm 0.0010$ 
in good agreement with results from the literature \cite{grassberger}. 
For completely uncorrelated random site sequences we have obtained
a significantly lower spreading temperature of  $T_s/T_c=0.9075 \pm 0.0010$. Up to our knowledge
there are no published data for DS in the case of  a random site sequence. (In Refs. \cite{costa,caer}
regular site order was used. Moreover the accuracy would not have been high enough to distinguish the 
different cases.)

From our results we conclude that the stationary state of DS is very sensitive to changes in the details
of the Monte-Carlo procedure even if they do not influence the stationary state of a single replica. 
For the ferromagnetic Glauber Ising model in three dimensions a change 
of the site order only leads to a shift of the spreading temperature $T_s$. For more complicated
systems it appears to be possible, however, that changing the site order leads to qualitative
changes of DS as was found for the change from sequential to parallel updates \cite{nobre}.
Investigations in this direction are in progress.

This work was supported in part by the DFG under grant
number  SFB393 and by the NSF under grant number 
DMR-95-10185.



\begin{references}
\bibitem{rn} In the case of Glauber dynamics
    the random number determines whether the spin is flipped, while in the heat-bath case it 
    determines whether it is set to $\pm 1$.
\end{references}
\end{document}